\documentclass[aps,twocolumn,showpacs,amsmath,amssymb,nofootinbib]{revtex4-1}
\usepackage{graphicx}
\usepackage{dcolumn}
\usepackage{bm}
\usepackage{color}
\usepackage{graphicx}
\usepackage{epsf}
\usepackage{wrapfig}
\usepackage{epsfig}
\usepackage{sublabel}
\usepackage{epsfig}

\def\Npart{{N_\text{part}}}

\def\b0{{\mbox{\boldmath$0$}}}

\def\Vec#1{\mbox{\boldmath $#1$}}

\def\beq{\begin{equation}}
\def\eeq{\end{equation}}

\def\beqy{\begin{eqnarray}}
\def\eeqy{\end{eqnarray}}

\def \b #1{ {\bf #1}}

\def \b #1{ {\bf #1}}

\def \b #1{ {\bf #1}}

\newcommand{\be}{\begin{eqnarray}}
\newcommand{\ee}{\end{eqnarray}}

\mathchardef\bbkappa="7114
\mathchardef\bbrho="711A
\mathchardef\bbsigma="711B
\mathchardef\bbtau="711C
\mathchardef\bbvarrho="7125
\mathchardef\bbvarsigma="7126
\mathchardef\bbxi="7118
\font\tenbifull=cmmib10 scaled 1200 
\font\tenbimed=cmmib9
\font\tenbismall=cmmib7
\begin{document} 
\title{Initial state anisotropies and their uncertainties in ultrarelativistic 
  heavy-ion collisions from the Monte Carlo Glauber model}
\author{M. Alvioli$^a$}
\email{Massimiliano.Alvioli@pg.infn.it}
\author{H.~Holopainen$^{b,c,d}$} 
\email{holopainen@fias.uni-frankfurt.de}
\author{K.~J.~Eskola$^{b,c}$}
\email{kari.eskola@phys.jyu.fi}
\author{M. Strikman$^e$}
\email{strikman@phys.psu.edu}
\affiliation{$^a$ ECT*, European Centre for Theoretical Studies in Nuclear Physics and Related Areas, 
  Strada delle Tabarelle 286, I-38123 Villazzano (TN) Italy}
\affiliation{$^b$ Department of Physics, P.O.Box 35, FI-40014 University of Jyv\"askyl\"a, Finland}
\affiliation{$^c$ Helsinki Institute of Physics, P.O.Box 64, FI-00014 University of Helsinki, Finland}
\affiliation{$^d$ Frankfurt Institute for Advanced Studies, Ruth-Moufang-Strasse 1, D-60438 Frankfurt am Main, Germany}
\affiliation{$^e$ The Pennsylvania State University, 104 Davey Lab, University Park, Pennsylvania 16803, USA}
\begin{abstract}
In hydrodynamical modeling of heavy-ion collisions, the initial-state spatial anisotropies are translated
into momentum anisotropies of the final-state particle distributions. Thus, understanding the origin of the 
initial-state anisotropies and their uncertainties is important before extracting  specific QCD matter 
properties, such as viscosity, from the experimental data. In this work we review the wounded nucleon 
approach based on the Monte Carlo Glauber model, charting in particular the uncertainties arising from 
modeling of the nucleon-nucleon interactions between the colliding nucleon pairs and nucleon-nucleon 
correlations inside the colliding nuclei. 
We discuss the differences between the black disk model and a probabilistic profile function approach
for the inelastic nucleon-nucleon interactions, and investigate the influence of initial-state
correlations using state-of-the-art modeling of these.
\end{abstract} 
\maketitle 
\section{Introduction}
In ultrarelativistic heavy ion collisions performed at Relativistic Heavy Ion Collider
(RHIC) and Large Hadron Collider (LHC) elliptic flow --
the second Fourier coefficient $v_2$ that quantifies the azimuthal anisotropy in the
measured particle distributions -- has been found to be large \cite{Adcox:2002ms,
Ackermann:2000tr,Aamodt:2010pa}. The appearance of a significant $v_2$ is a clear signature 
of pressure formation in the system. It is also consistent with predictions from relativistic 
hydrodynamics \cite{Huovinen:2001cy,Ollitrault:1992bk}: At non zero impact parameter the 
overlap area of the colliding nuclei is eccentric in the transverse coordinate plane. 
This spatial anisotropy  is translated first into a flow-velocity anisotropy during the 
hydrodynamical evolution and, finally, at the decoupling of the system, into a measurable 
momentum-anisotropy of final-state particle distributions.

Recently it was found out that geometrical fluctuations in the nucleon transverse
positions generate initial state anisotropies \cite{Alver:2010gr} also for odd
harmonics $v_n$. Nowadays, the experiments at RHIC and LHC have been able to 
measure non zero flow coefficients up to $v_6$ \cite{Adare:2011tg,Sorensen:2011fb,
:2011vk,collaboration:2011hfa}. In recent years, there have also been considerable 
developments in event-by-event hydrodynamical modeling \cite{Andrade:2006yh,Andrade:2008xh,
Petersen:2009vx,Werner:2010aa,Holopainen:2010gz,Schenke:2010rr}, thanks to which it has
become possible to study the higher harmonics \cite{Qin:2010pf,Qiu:2011iv,Schenke:2011bn} 
and $v_1$ at mid rapidity \cite{Teaney:2010vd,Luzum:2010fb,Gardim:2011qn}. 
Studies based on hydrodynamics have shown that all initial-state anisotropies are 
transferred to the final measurable flow values in a similar way as eccentricity 
is translated to elliptic flow.

In the past few years much effort has been put on developing viscous hydrodynamics
\cite{Romatschke:2007mq,Dusling:2007gi,Luzum:2008cw,Luzum:2009sb,
Song:2007fn,Song:2007ux,Song:2008si,Heinz:2009cv,Shen:2010uy,Song:2010mg,
Song:2010aq,Song:2011hk,Song:2011qa,Niemi:2011ix}
and it has turned out that the elliptic flow and especially the higher harmonics are sensitive
to the (shear) viscosity $\eta$ \cite{Alver:2010dn,Schenke:2010rr,Schenke:2011bn,Staig:2011wj}.
Since the final flow observables closely reflect the initial-state spatial anisotropies,
the uncertainties in the assumed (or computed) initial state are transferred to the
computed final-state flow observables. In the studies where one has tried to estimate the 
shear viscosity-to-entropy ratio, $\eta/s$, using elliptic flow measurements, one has seen that
the initial eccentricity differences between different initial state models lead to
large uncertainties in the extracted value of $\eta/s$ \cite{Luzum:2008cw,Song:2010mg}.
For determining the shear viscosity from the flow measurements, it is very important to 
chart all the relevant uncertainties in the computation of the initial asymmetries.

The Glauber model \cite{glauber} is usually a key element in computing the initial states for  
hydrodynamical modeling of ultrarelativistic heavy-ion collisions. Some years back, most 
hydrodynamical calculations assumed smooth initial states where the (energy or entropy) densities 
were assumed to scale with the density of binary collisions or wounded nucleons computed from the 
optical Glauber model; see, e.g. \cite{Kolb:2001qz}. Now that the importance of the initial density 
fluctuations has been realized, Monte Carlo Glauber (MCG) modeling has become more frequently used. 
So far the black disk (hard-sphere) modeling of the nucleon-nucleon ($NN$) interactions has been the 
standard choice \cite{Alver:2008aq,Alver:2008zza,Hirano:2009ah} in these studies, although also more 
involved probabilistic ways to model the $NN$ interactions have been known for a long time 
\cite{glauber,Wang:1990qp,Pi:1992ug,Glauber:1970jm}.

For MCG modeling, one needs the nucleon configurations inside the colliding nuclei as an 
input for each event. In the simplest approximation the nucleons are assumed point-like and 
the nucleon positions are obtained by just randomly sampling the Woods-Saxon number density 
distribution. However, to improve the modeling, $NN$ correlations should also be considered, as 
suggested in Refs. \cite{Alvioli:2009ab,Alvioli:2010yk,Alvioli:2008rw,Alvioli:2009iw,
CiofidegliAtti:2011fh}. 
Experimental evidence for the two-body $NN$ correlations generating high-momentum components 
of the nuclear wave function \cite{Alvioli:2005cz,Alvioli:2007zz}, are discussed in  
Refs.~\cite{Tang:2002ww,Shneor:2007tu,Subedi:2008zz}. Furthermore, recent studies in 
Ref.~\cite{Broniowski:2010jd} have shown that the effects from including more realistic 
$NN$ correlations are not negligible e.g. in the generated initial eccentricity.

In this paper we study for Au-Au collision at RHIC energies, two different uncertainties 
in computing the initial state asymmetries from the MCG model:
one related to the modeling of the inelastic $NN$ collisions between nucleons 
from different nuclei, and one related to the $NN$ correlations in the nucleon configurations in 
each of the colliding nuclei. The paper is organized as follows: In Sec.~\ref{sec: nucl conf} we 
outline how the initial nucleon configurations can be obtained taking the $NN$ correlations into 
account. After that, in Sec.~\ref{sec: interaction}, we present two different ways to model the 
inelastic $NN$ primary collisions. The studied initial-state anisotropies are defined in 
Sec.~\ref{sec: anisotropy}, and results are presented in Sec.~\ref{sec: results}. 
Finally, the conclusions are drawn in Sec.~\ref{sec: conclusions}.

\section{Monte Carlo Glauber model framework}

\subsection{Nucleon configurations}
\label{sec: nucl conf}

The initial state of a nucleus in the MCG calculations is usually 
taken as a collection of particles distributed according to a probability distribution 
given by the corresponding (Woods-Saxon) number density distribution measured in electron scattering
experiments. Given the complexity of the nuclear many-body problem, the effects of spatial, spin and
isospin dependent correlations among the nucleons are usually overlooked and the nucleons are 
positioned randomly for each of the simulated events.

Recently, in Ref.~\cite{Alvioli:2009ab} it was shown how such an approach can be modified by including 
initial states, which are prepared in advance, in the commonly used computer codes. Also, a method to 
produce such configurations was introduced. The method is based on the notion of a nuclear wave function 
$\psi$, which contains the nucleonic degrees of freedom and which is used to iteratively modify the positions 
of randomly distributed nucleons using the Metropolis method so that the final positions correspond to the 
probability density given by $|\psi|^2$. The method is constructed to reproduce the same nucleon number-density 
distribution as the usual one and, in addition, to reproduce the basic features of the two-nucleon density in 
the presence of the $NN$ correlations. The model wave function is taken in the form
\beq
\psi(\Vec{x}_1,...,\Vec{x}_A)=\prod^A_{i<j}\hat{f}_{ij}\,\phi(\Vec{x}_1,...,\Vec{x}_A)\,,
\eeq
where $\phi$ is the uncorrelated wave function and $\hat{f}_{ij}$ are correlation
operators; here, $\Vec{x}_i$ denotes the position, spin and isospin projection of the
$i$-th nucleon. The correlation operator contains a detailed spin-isospin dependence,
which is as follows:
\beq
\begin{split}
\label{massi1}
\hat{f}_{ij}\,=&\,f^c(r_{ij}) +f^\sigma(r_{ij})\Vec{\sigma}_i\cdot\Vec{\sigma}_j 
+f^\tau(r_{ij})\Vec{\tau}_i\cdot\Vec{\tau}_j  \\
&+f^{\sigma\tau}(r_{ij})\Vec{\sigma}_i\cdot\Vec{\sigma}_j\Vec{\tau}_i\cdot\Vec{\tau}_j 
+f^{t}(r_{ij})\hat{\Vec{S}}_{ij} \\
&+f^{t\tau}(r_{ij})\hat{\Vec{S}}_{ij}\Vec{\tau}_i\cdot\Vec{\tau}_j +...\,,
\end{split}
\eeq
where $r_{ij}$ is the relative distance between nucleons $i$ and $j$, $\Vec{\sigma_i}$
and $\Vec{\tau_i}$ are the Pauli spin and isospin operators, respectively, and
$\Vec{\hat{S}_{ij}}$ is the so-called tensor operator (see e.g.
Ref.~\cite{Alvioli:2005cz}) which depends on the spin and spatial variables of
nucleons $i$ and $j$. In the most general case, the sum extends over a number of
channels that are
the same as the one appearing in modern nucleon-nucleon potentials used to successfully 
describe a variety of properties of light and medium-heavy nuclei within different 
\textit{ab initi}o approaches. When all the correlation functions in Eq. (\ref{massi1}), 
except the first one [$f^c(r)$], vanish, we have the central correlation case.

In this paper we will consider configurations produced both with the central correlations 
only (denoted in the following by \textit{central}) and with a full set of the six correlation 
functions of Eq.~(\ref{massi1}) (denoted by \textit{full}), including the tensor operator. 
These correlation functions \cite{Alvioli:2005cz} were developed by variational method for 
nuclei lighter than $^{197}$Au considered in this paper; nevertheless, they represent the 
best approximation that we can offer, since no corresponding calculations exist for heavy 
nuclei. More specifically, we used correlation functions obtained for $^{40}$Ca. Many
recent theoretical works support the universality of $NN$ correlations in nuclei 
\cite{Alvioli:2011aa,Feldmeier:2011qy,Vanhalst:2011es}. We justify
the use of $^{40}$Ca correlation functions by the observation that they differ very little 
from those obtained for $^{16}$O; $^{40}$Ca is a large enough nucleus to neglect additional 
differences with $^{197}$Au due to $A$ dependence in this context, as well as differences arising 
from the fact that correlation functions were obtained for doubly magic, symmetric nuclei such 
as $^{16}$O and $^{40}$Ca.
Moreover, as shown in 
Ref.~\cite{Alvioli:2010yk}, the two-body densities resulting from configurations obtained 
using these correlations are clearly more realistic than the completely uncorrelated ones.

The method developed in Ref.~\cite{Alvioli:2009ab} allows in principle to deal with any set 
of correlations; nevertheless, although the state dependent correlations introduce an extra 
computational effort due to their non commutative nature, they can be treated up to a certain 
degree. In this paper we have used configurations including two-body full correlations, and 
configurations of three-body clusters surrounding each of the nucleons, induced by full 
correlations. Genuine three-body correlations will also be  discussed in the following. 

\subsection{Modeling the inelastic interactions}
\label{sec: interaction}

We work in the Glauber model framework \cite{glauber}, neglecting the effects of inelastic 
diffraction that lead to fluctuations of the strength of the $NN$ interactions
\cite{Blaettel:1993ah,Baym:1995cz}. To generate the inelastic $NN$ collisions of interest here, 
we use the following two different approximations for deciding whether a collision between the 
nucleons $i$ and $j$ from different nuclei takes place: 

\noindent
\textbullet~
\textit{Black disk} approximation, used recently, e.g., in Ref. \cite{Holopainen:2010gz}, where 
one assumes the two nucleons to interact inelastically with a probability one if their transverse 
separation $b_{ij}$ is within a radius defined by the inelastic $NN$ cross section $\sigma^{in}_{NN}$,
\beq
\label{bdisk}
\Vec{b}^2_{ij}\le \frac{\sigma^{in}_{NN}}{\pi}\,;
\eeq

\noindent
\textbullet~
\textit{Profile function approach}, where the probability of an inelastic interaction between the 
nucleons $i$ and $j$ is given by
\beq
\label{intprob}
P(\Vec{b}_{ij})\,=\,1\,-\,\left|1-\Gamma(\Vec{b}_{ij})\right|^2\,,
\eeq
and where the profile function $\Gamma$ is expressed in terms of the total and elastic $NN$ cross sections 
as follows:
\beq
\label{gamma}
\Gamma(\Vec{b}_{ij})\,=\,\frac{\sigma^{tot}_{NN}}{4 \pi B}\,e^{-b^2_{ij}/(2\,B)}\,,
\eeq
with $B=(\sigma^{tot}_{NN})^2/(16\pi\sigma^{el}_{NN})$.

The probability distribution $P(\Vec{b}_{ij})$ can be derived in the Born approximation of the 
potential-scattering formalism, by parametrizing the $NN$ elastic scattering amplitude as 
\cite{glauber,Blaettel:1993ah,Alvioli:2008rw}
\beq
\label{ampl}
f(\Vec{q})\,=\,\frac{C(i\,+\,\alpha)k}{4\pi}\,e^{-\frac{1}{2}B q^2}\,,
\eeq
where $C$ and $B$ are constants to be determined, $k\approx E$ and $\alpha\approx 0$ for ultra-relativistic 
energies. For the small-angle scatterings of interest here, the vector \textbf{q} corresponds to the 
difference between the incoming and scattered wave vectors in the transverse plane, and $q=|\textbf{q}|$ 
relates to the scattering angle as $q\approx k\theta$. We fix $C$ to the measured $\sigma^{tot}_{NN}$ on 
the basis of the optical theorem, 
\beq
\label{opt_th}
\sigma^{tot}_{NN} = \frac{4\pi}{k} Im[f(\Vec{0})].
\eeq
The profile function in Eq. (\ref{gamma}) is obtained as a Fourier transform,
\beq
\label{gammb}
\Gamma(\Vec{b}_{ij})\,=\,\frac{1}{2\pi i k}\int d^2\Vec{q}\,e^{-i\,\Vec{q}\cdot\Vec{b}_{ij}}\,f(\Vec{q})\,,
\eeq
and the elastic $NN$ cross section from this as
\beq
\label{elas}
\sigma^{el}_{NN}\,=\,\int d^2\Vec{b}_{ij}\,\left|\Gamma(\Vec{b}_{ij})\right|^2.
\eeq
Thus, we have 
\beqy
\label{ssel}
\sigma^{el}_{NN}&=&\frac{\left(\sigma^{tot}_{NN}\right)^2}{16 \pi B}\,,\\
\label{stot}
\sigma^{tot}_{NN}&=&\sigma^{el}_{NN}\,+\,\sigma^{in}_{NN}\,,
\eeqy
and $B$ can be fixed on the basis of measured cross sections. For the current set-up at $\sqrt {s_{NN}}=200$~GeV, 
we take $B=14$~GeV$^{-2}$ and $\sigma^{tot}_{NN}= 52$~mb, which correspond to $\sigma^{in}_{NN}= 42$~mb and 
$\sigma^{el}_{NN}= 9.9$~mb. 

Finally,  using
Eqs.~(\ref{opt_th}), (\ref{elas}) and (\ref{stot}), we arrive at the probability function of Eq.~(\ref{intprob}), 
whose integral over the transverse separation gives the inelastic $NN$ cross section:
\beqy
\sigma^{in}_{NN}&=&\int d^2\Vec{b}_{ij}\,\left(2\, Re\,\Gamma(\Vec{b}_{ij})\,-\,\left|\Gamma(\Vec{b}_{ij})\right|^2\right)\,\nonumber\\
\label{inel}
&=&\int d^2\Vec{b}_{ij}\,\left(1\,-\,\left|1\,-\,\Gamma(\Vec{b}_{ij})\right|^2\right)\,.
\eeqy

\subsection{Spatial asymmetries and their fluctuations}
\label{sec: anisotropy}

In this section we define the initial state anisotropies, which are studied in this work.
We focus on the first three harmonics $n=1,2,3$: dipole asymmetry, eccentricity, and
triangularity. Higher harmonics are left out since they are more complicated due to
the fact that they can be mainly originating from the lower harmonics
\cite{Borghini:2005kd}. Also, from the experimental results we know that the second and third
harmonics are the largest ones.

We calculate the asymmetries from the wounded nucleon positions which are obtained
from the MCG model. In the following, the angle brackets denote an average over wounded,
or participant, nucleons. The asymmetries are defined as
\begin{equation}
  \epsilon_n =  -\frac{ \langle w(r) \cos (n(\phi - \psi_n) \rangle }{ \langle w(r) \rangle },
\end{equation}
where $w(r)$ is a weight and $\psi_n$ is an orientation angle that is obtained as
\begin{equation}
  \psi_n = \frac{1}{n} \arctan \frac{\langle w(r) \sin (n\phi) \rangle }{ \langle w(r) \cos (n\phi) \rangle } + \frac{\pi }{n},
\end{equation}
where $\arctan$ is always placed in the correct quadrant. These quantities are always
calculated in the center-of-mass of the wounded nucleon system. We choose to use $w(r) = r^3$
for $n=1$, $w(r) = r^2$ for $n=2$, and $w(r) = r^3$ for $n=3$ \cite{Teaney:2010vd}.

We also study the fluctuations of the initial-state asymmetries since 
the current flow analysis methods are sensitive to the fluctuations of the flow 
coefficients \cite{Ollitrault:2009ie}. Since final flow values reflect the 
initial-state asymmetries, the fluctuations of the flow coefficients should follow the 
initial-state fluctuations. We define the fluctuations of the anisotropies as
\begin{equation}
  \Delta \epsilon_n = \sqrt{\frac{\sum (\epsilon_n^i - \langle \epsilon_n \rangle)^2 }{N} },
\end{equation} 
where $\epsilon_n^i$ is the asymmetry in event $i$ and $\langle \epsilon_n  \rangle$
denotes the average over $N$ events.

\section{Results}
\label{sec: results}

\subsection{Number of wounded nucleons and binary collisions}

\begin{figure}[h]
  \includegraphics[width=9.0cm]{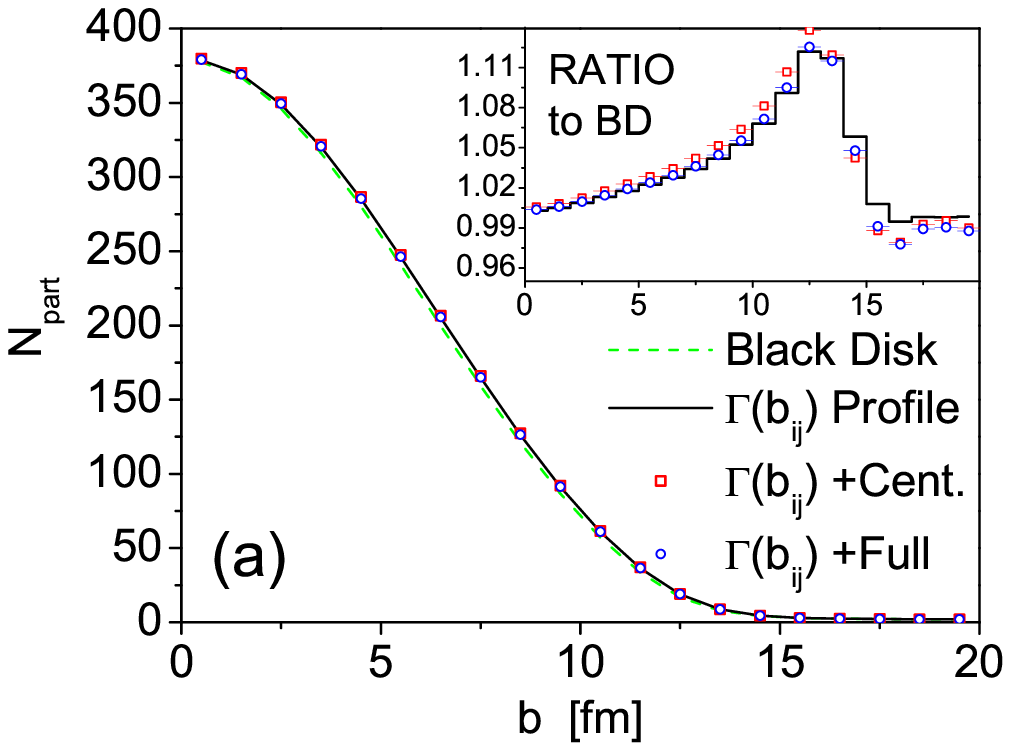}
  \includegraphics[width=9.0cm]{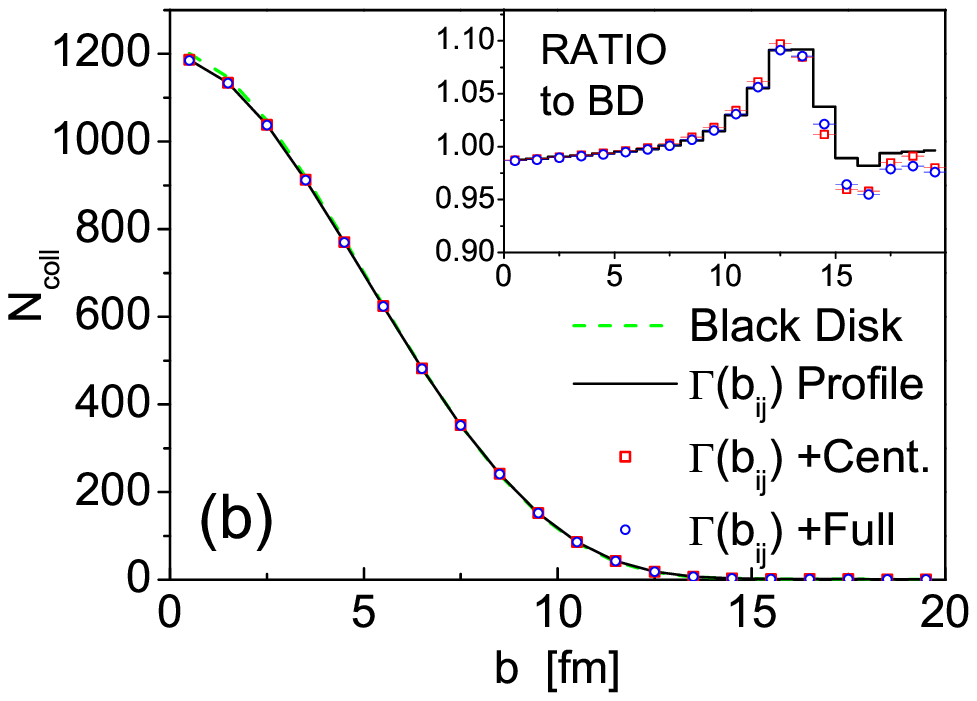}
  \caption{\protect\small (Color online) Number of wounded nucleons (a)
    and binary collisions (b)
    as a function of impact parameter with different interaction models and initial-state
    configurations. The results shown correspond to 
    the black disk approximation with uncorrelated configurations (green dashed line); 
    the profile function approach $\Gamma(b_{ij})$ with uncorrelated configurations (black solid line);
    the profile function approach with central $NN$ correlations (red squares); and 
    the profile function approach with full $NN$ configurations (blue circles).
    The insets show the ratios of the last three cases to the black disk + uncorrelated one, with
    corresponding notations.}
  \label{fig: wounded}
\end{figure}

First we look how the interaction models and $NN$ correlations affect the number of the
participants ($N_{\rm part}$) and binary collisions ($N_{\rm coll}$). In Fig.~\ref{fig: wounded}
we have plotted these as a function of the impact parameter. One can see that in central collisions 
the results are the same with both interaction models, but when impact parameter becomes large, 
$b=10-15$~fm, i.e., when only the edges of the nuclei collide and when the number of participants 
and binary collisions are of the same order, the difference between models is over 5~\%. This holds 
both for wounded nucleons and binary collisions. From Fig.~\ref{fig: wounded} it can be seen that 
the effects of $NN$ correlations on these quantities are very small.

\subsection{Effects of the interaction model on $\epsilon_n$ and their fluctuations}

\begin{figure}[t]
  \includegraphics[width=9.0cm]{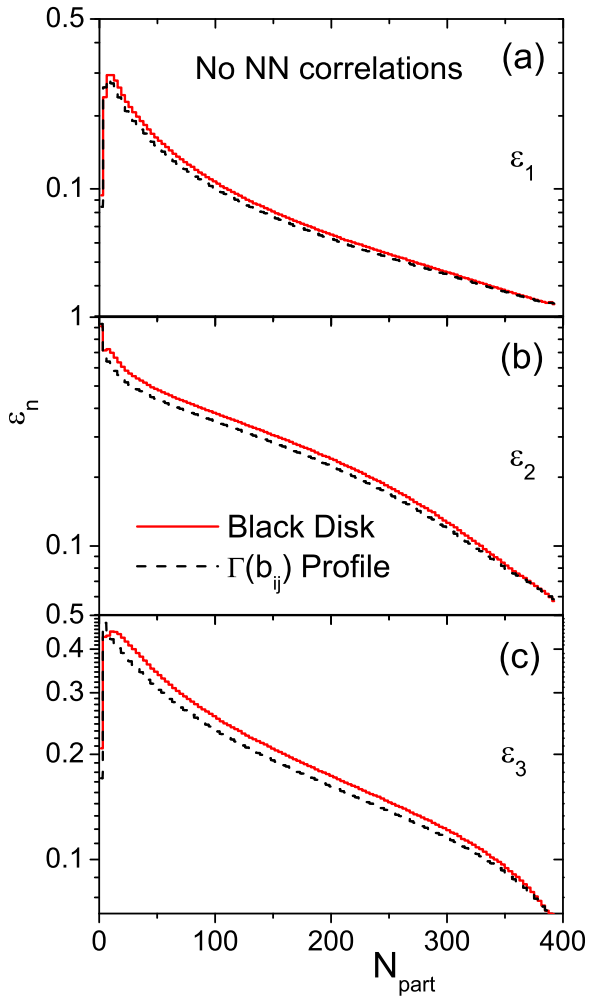}
  \caption{\protect\small (Color online) First three harmonics $\epsilon_n$ 
    as a function of the number of participants. Results shown are for the two different
    interaction models. In these plots no $NN$ correlations are taken into account.}
  \label{fig: interaction eps}
\end{figure}

Next, we consider the two different interaction models discussed 
in Sec.~\ref{sec: interaction} and present their effects on the anisotropies defined above.
In these calculations the $NN$ correlations are neglected for clarity; the
effects of these correlations will be studied in the next section.
In Fig.~\ref{fig: interaction eps} we have plotted the anisotropies $\epsilon_1$, $\epsilon_2$, and
$\epsilon_3$ as a function of $N_{\text{part}}$ using the black disk and profile function approaches
to model the inelastic $NN$ collisions. The black disk approximation results in larger dipole asymmetry 
$\epsilon_1$, eccentricity $\epsilon_2$, and triangularity $\epsilon_3$.
In all cases, the obtained asymmetries in the profile function approach are slightly ($\lesssim 10$ ~\%)
smaller than in those in the black disk case. Apparently, toward peripheral collisions, we have more 
fluctuations and a less-well-defined shape with the profile function approach, hence a smaller
$\epsilon_n$.

The difference between these two cases is negligible in most central collisions for
every $n$. When moving toward peripheral collisions, the results start to deviate from 
each other and the largest difference is approximately on the order of 10\%. In central
collisions most of the nucleons experience several collisions and, thus, the details 
of the collision model are not very important. In peripheral collisions, however, more and more 
of the nucleons collide only once or twice, meaning that the interaction model details 
start to play a role. One must also remember that as we saw from Fig.~\ref{fig: wounded},
the events with the same impact parameter are mapped to slightly smaller $\Npart$ values with
the black disk interaction model.

\begin{figure}[h]
  \includegraphics[width=9.0cm]{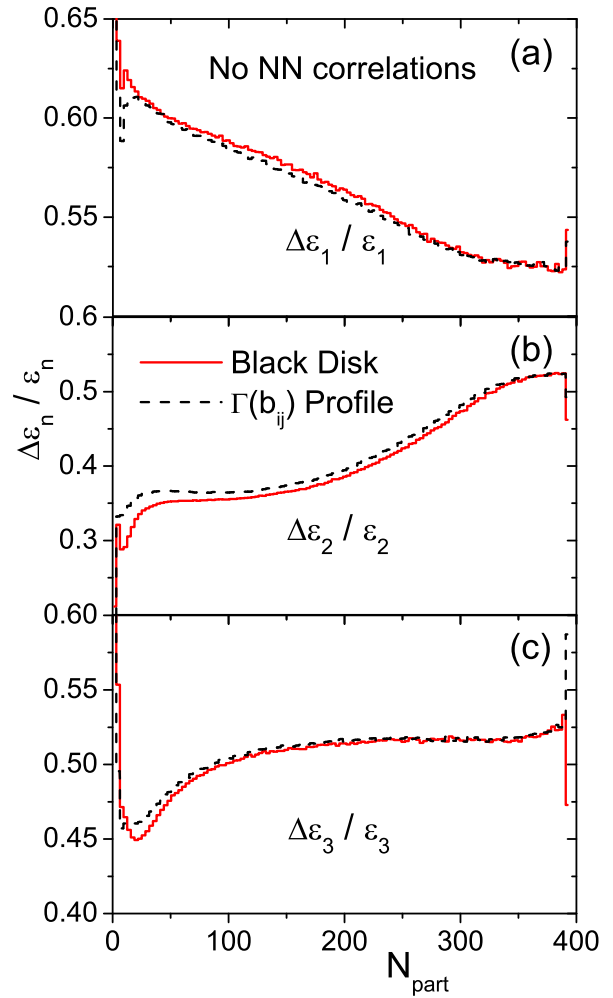}
  \caption{\protect\small (Color online) Relative fluctuations of the first three
    harmonics $\epsilon_n$ as a function of number of participants. Results are shown 
    with two different interaction models. In these plots no $NN$ correlations are taken 
    into account.
  }
  \label{fig: interaction sigma}
\end{figure}

In Fig.~\ref{fig: interaction sigma} we plot the relative fluctuations of the initial-state
anisotropies. Here the order of the curves is opposite compared to the $\epsilon_n$ when 
$n=2$, $3$. This means that at least partly the difference in the relative fluctuations
is explained simply by the fact that the absolute value is slightly higher in the
case where the relative fluctuations are smaller. For $\epsilon_1$ the relative fluctuations 
are larger with black disk interaction, indicating that then also the absolute fluctuations 
are larger with black disk than profile interaction.

The relative fluctuations of $\epsilon_1$ show a decreasing trend when approaching central
collisions but the impact-parameter dependence is relatively weak.
On the other hand, the relative fluctuations of $\epsilon_2$ clearly depend on the impact 
parameter. In central collisions the fluctuations are large since the system is
azimuthally very symmetric, and toward peripheral collisions the fluctuations decrease,
since the collision area becomes clearly eccentric. Triangularity $\epsilon_3$ has yet 
another behavior: its relative fluctuations stay approximately constant when $\Npart > 100$.
Since the triangularity is created purely by the fluctuations in the positions of wounded 
nucleons, we can expect that the relative fluctuations have no centrality dependence. 
All in all, $\Delta\epsilon_1/\epsilon_1$ and $\Delta \epsilon_3/\epsilon_3$ have only a
mild centrality dependence, while $\Delta \epsilon_2/\epsilon_2$ exhibits a clear dependence 
on the centrality.
In addition, we conclude that for these relative fluctuations, the $NN$ interaction model uncertainties 
do not play a major role.

\subsection{Effects of the NN correlations on $\epsilon_n$ and their fluctuations}

\begin{figure}[h]
  \includegraphics[width=9.0cm]{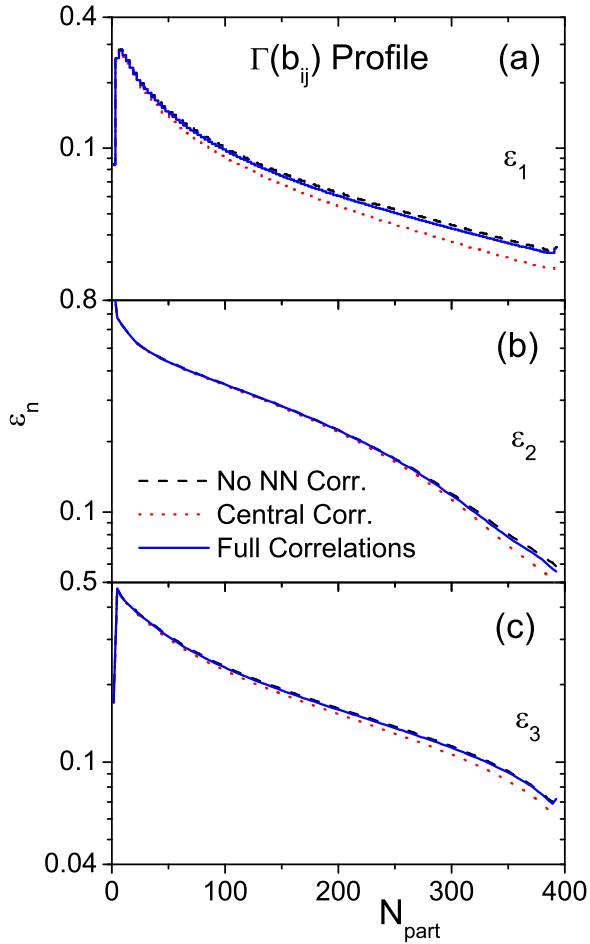}
  \caption{\protect\small (Color online) First three harmonics $\epsilon_n$ as a function
    of number of participants. Results are shown with different $NN$ correlations and
    using the $\Gamma(b_{ij})$ collision profile function approach.}
  \label{fig: correlation eps}
\end{figure}

Next, we investigate the effects of different models for the initial state
$NN$ correlations. We have chosen to use the $\Gamma(b_{ij})$ collision
profile function in these calculations. In Fig.~\ref{fig: correlation eps}
we have plotted the obtained anisotropies with three different $NN$ correlations: no 
correlations at all, only central correlations and full correlations. From all panels 
we see that no correlations and full correlations are very close to each other, but 
the central correlations have a smaller anisotropy. 

In all cases the difference between central correlations and the two other cases is
largest at central collisions and it gets smaller toward peripheral collision.
Nucleon correlations are most important in the middle of the nucleus since the nucleon
density is largest there. Thus, in the central collisions, where most of the
wounded nucleons come from the middle of the nuclei, the effect on anisotropies is
largest. When moving toward the peripheral collisions, the relative amount of wounded
nucleons coming from the edges of nuclei is increasing. Thus, in very peripheral
collisions where only the edges are overlapping, the effect disappears.

We must bear in mind that central correlations have only repulsive character; their
effect can be mimicked by an ad-hoc exclusion radius, as shown in
Ref.~\cite{Broniowski:2010jd}.
Full state-dependent correlations, instead, have a complex structure, which causes
nucleons to be at a given average distance in the nucleus and results in nuclear binding 
and its saturation with increasing $A$. Partially including these full correlations and 
\begin{figure}[!ht]
  \centering
  \includegraphics[width=9.0cm]{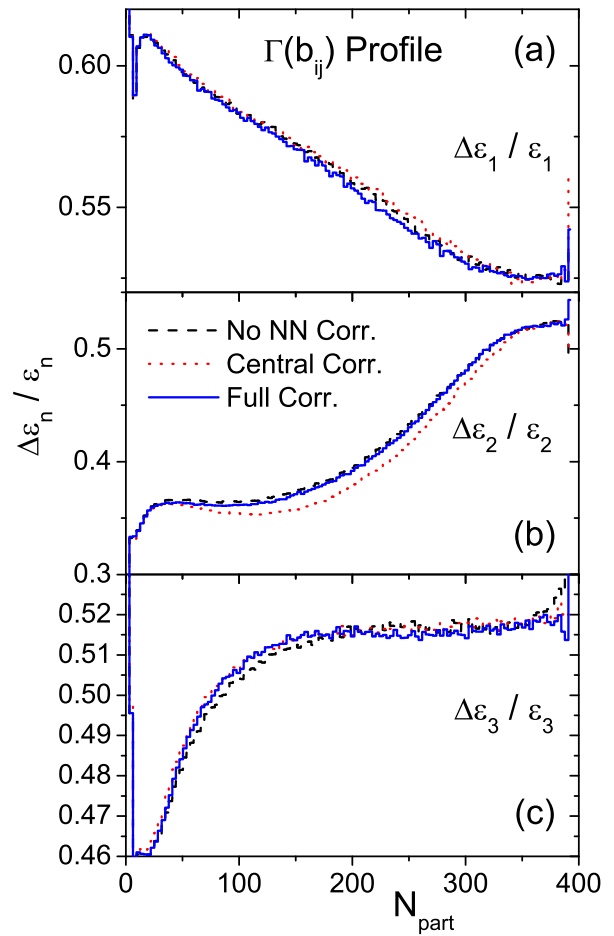}
  \caption{\protect\small (Color online) Relative fluctuations of the first three
    harmonics $\epsilon_n$ as a function of the number of participants. Results are shown
    with different $NN$ correlations and using the $\Gamma(b_{ij})$ collision profile approach.}
  \label{fig: correlation sigma}
\end{figure}
disregarding three-body repulsion, as it is explained below, produces the net result 
of working  in the opposite direction of repulsion, for the considered quantities.

Next, in Fig.~\ref{fig: correlation sigma}, we plot the relative fluctuations of anisotropies 
with different initial state $NN$ correlations. We can see that inclusion of full correlations 
brings the results back toward the uncorrelated case; this is less evident for 
$\Delta\epsilon_3/\epsilon_3$. However, now the difference is small in the central collisions 
and it is largest in the semi peripheral collisions. The difference again vanishes at the most 
peripheral collisions.

The results obtained with realistic configurations deserve some discussion.
The production of configurations with full realistic correlation functions
differs substantially from the central correlations case, since in the realistic
description there are several spatially-dependent correlation functions complemented
with spin- and isospin-dependent operators, as shown in Eq.~\eqref{massi1}.
At this stage, we have performed a truncation of the chains induced by
realistic correlations at the level of three-particle chains; in principle,
due to the non commutativity of the operators in Eq.~\eqref{massi1}, the chains
are $A$-body operators. Moreover, we restricted our calculations to three particles
which are within a given radius from the active particle in the Metropolis
algorithm. It should be stressed that these kinds of simplifications are not
\begin{figure}[!ht]
  \centering
  \includegraphics[width=9.0cm]{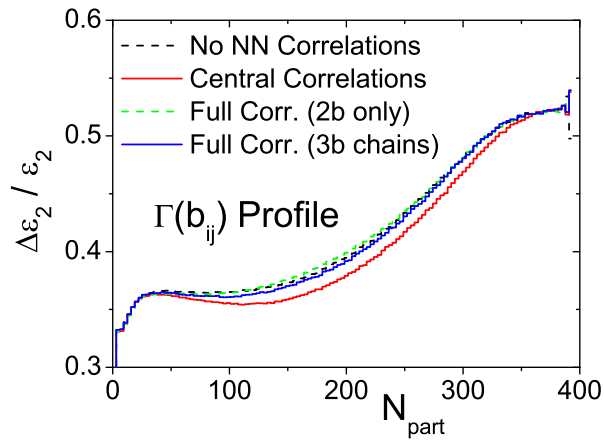}
  \caption{\protect\small (Color online) The effect of different models of realistic correlations
    on the relative fluctations of eccentricity; see text for explanation.}
  \label{fig: full correlations}
\end{figure}
suggested by specific physics arguments but are rather dictated by the enormously
increasing computing time. As a result, the truncation obviously induces some amount 
of uncertainty in our results. The overall trend, that the full correlations seem to
bring the results back toward the no correlation case, is nevertheless what we wish to 
emphasize here.   

In order to illustrate the effect of the truncation, we have repeated the calculation 
with configurations including only two-body, realistically correlated clusters (2b only), 
and compared with the three-body calculation outlined above (3b chains). The results for 
$\Delta\epsilon_2/\epsilon_2$ are compared with the corresponding uncorrelated and central 
correlation case in Fig.~\ref{fig: full correlations}. Interestingly, it can be seen that 
the full correlations with the 2 body chains cause an effect into the opposite direction 
than the more advanced 3b chains do. 
Although beyond the scope of this paper, it would be interesting to study how sensitive the 
considered anisotropies, and also the higher moments of participant matter distribution, are 
to the higher-order chains of two-body state-dependent correlations and to the genuine three-body 
correlations which we have not included here. The outcome of such a study is, however, difficult 
to predict, due to the complicated interplay of attraction and repulsion between the three 
particles in different spin and isospin states. 

\section{Conclusions}
\label{sec: conclusions}

In this paper we have charted some of the uncertainties in the computation of the initial state anisotropies
from the Monte Carlo Glauber model. We used two different ways of modeling the inelastic interactions between
the colliding nucleons. The difference between these two cases gives us an estimate about the uncertainties 
related to this part of the model: in central collisions the details of the interaction model play a minor 
role, but in the peripheral collisions such details can cause uncertainties up to 10\% in the first three 
harmonics $\epsilon_1$, $\epsilon_2$ and $\epsilon_3$. We also checked that with these two interaction models 
the difference in the number of wounded nucleons and binary collisions remains small in central collisions, 
but at impact parameters 10-15 fm the difference can be around 10\%. We also note that during the writing 
process of this article, a similar nucleon interaction model study was released in Ref.~\cite{Rybczynski:2011wv}. 
The main differences to our study are the different form of the elastic $NN$ scattering amplitude as well as the 
treatment of the $NN$ correlations.

We also presented a study of the effects of $NN$ correlations with an update of correlated
configurations and extended discussion as compared with the previous published papers on
this subject. We confirmed that the inclusion of centrally correlated nucleon
configurations produce the effects to eccentricity and its relative variance as was
claimed by Ref.~\cite{Broniowski:2010jd}. 
As a new result, we observed that the inclusion of realistically correlated configurations (two-body full 
correlations, three-body chains) seems to essentially cancel this effect and bring the results back close to 
the no correlations case. The effect is similar for dipole asymmetry and triangularity as for eccentricity.
However, we also showed that there are still uncertainties caused by the truncation done in the nucleon 
configuration calculation with full correlations and we expect three-body correlations to play a role.

In this paper we studied two sources of additional uncertainty in the Monte Carlo model 
calculations for the initial state anisotropies. The uncertainty caused by the studied 
effects to these anisotropies was found to be maximally of the order of 10\%. 
Now that -- thanks to the recent developments in event-by-event hydrodynamics -- more 
precise comparisons of flow coefficients between the data and the theory are becoming 
possible, it is important to chart all the relevant uncertainties to this precision, so 
that the QCD matter properties could eventually be determined from the measured particle 
spectra and their azimuthal asymmetries.

\begin{acknowledgments} 
The work of M.A. is supported by the HadronPhysics2 project of the European Commission, 
Grant No. 227431.
The work of H.H. is supported by the national Graduate School of Particle and Nuclear
Physics in Finland and the Extreme Matter Institute (EMMI).
We gratefully acknowledge financial support from the Academy of Finland, K.J.E.'s Project
No. 133005.
We acknowledge CSC -- IT Center for Science in Espoo, Finland, for the allocation of
computational resources.
M.A. thanks the HPC-Europa2 consortium for a grant and the 
Department of Physics, University of Jyv\"askyl\"a, for hospitality during the
development of this work.
We thank T. Lappi for useful discussions.

\end{acknowledgments}

\end{document}